# Finite Phase-separation FRET II: Determination of domain size from simulated data

*Frederick A. Heberle [a]* and Gerald W. Feigenson [b]*

[a,b]Field of Biophysics, Cornell University, Ithaca, NY 14853

AUTHOR EMAIL ADDRESS: [a]fah2@cornell.edu, [b]gwf3@cornell.edu

*To whom correspondence should be addressed. Email: fah2@cornell.edu, Phone: (865) 576-8802.

**ABSTRACT**

We have developed a new model to describe FRET efficiency ($E_{FRET}$) between freely-diffusing membrane probes in phase-separated bilayers (Finite Phase-separation FRET, or FP-FRET), that in principle applies to any system where phase domain dimensions are larger than ~ $R_0$. Here we use Monte Carlo techniques to simulate $E_{FRET}$ for a range of probe partitioning behaviors and domain sizes, and then fit the simulated data to the FP-FRET model to recover simulation parameters. We find that FP-FRET can determine domain size to within 5% of simulated values for domain diameters up to ~ 5 $R_0$, and to within 15% for diameters up to ~ 20 $R_0$. We also investigated the performance of the model in cases where specific model assumptions are not valid.

# INTRODUCTION

For biological membranes, the concept of a "raft" describes the compartmentalizing of membrane components, based on the differential physical and chemical properties of the raft and its surrounding "sea". There is evidence that the controlled size of plasma membrane rafts is an important aspect of their functionality (1). The extent to which a given component (a transmembrane protein or a lipid) is compartmentalized can be expressed by its partition coefficient between the raft and non-raft domains (if there are only two types of domain). Binding of molecules to regions of a membrane, and rates of chemical reactions within a membrane might be controlled not only by the lipid composition of a raft, but also by its size.

In resting cells, rafts cannot be visualized by conventional light microscopy, an observation that sets an approximate upper limit of 200 nm on raft spatial dimensions. Though precise measurements on nanometer scales are experimentally challenging, estimates for raft dimensions are converging on the order of 2-20 nm (2). After stimulation (for example by an addition of an external cross-linker), an increase in raft size is frequently observed. In the most extreme cases, micron-sized domains can be directly visualized with fluorescence microscopy.

In parallel with cell studies, observations of nanometer-scale heterogeneity in 3-component lipid mixtures have stimulated the interest of researchers. A lively debate exists as to the thermodynamic origins of small domains, with some evidence pointing toward genuine first-order phase separation (3), some evidence supporting Ising-like critical phenomena (4,5). The dependence of domain size on lipid composition and temperature may yield important insights into the thermodynamic nature of these domains. Furthermore, there is an obvious connection between observations of rafts in cells and small domains in model membranes, with a growing consensus that the physical origin of raft phenomena is closely related to lipid phase separation.

Only a few techniques can measure the size of small membrane domains, including FRET (6,7), AFM (8), and the newly-emerging super-resolution techniques (9,10). Unique among these techniques, FRET between freely-diffusing lipid probes does not require a bilayer support, which can introduce strong interactions (11,12). This is an especially important consideration for nanodomains, where even small perturbations might tip the delicate balance of interaction energies and cause significant artifacts in size measurements.

In the companion paper, we developed a model (FP-FRET) to predict energy transfer efficiency in a bilayer composed of nanoscopic phase domains. Using simulated data, we report on the performance of the FP-FRET model for predicting the size of phase domains, and show that a global analysis of multiple donor-acceptor pairs can yield accurate estimates for domain diameters up to 20 times the Förster distance. We also investigate the reliability of recovered parameters when various physical assumptions of the model are not strictly met.

**MATERIALS AND METHODS**

Monte Carlo simulations of energy transfer efficiency

All simulation and data analysis was performed with Mathematica 7.0.1 (Wolfram Research, Champaign, IL). Energy transfer simulations for donor and acceptor fluorophores in phase-separated bilayers were performed by constructing random snapshots of probe configurations subject to the considerations described below.

*Bilayer parameters*

A key aspect of these studies is that simulations are based on a rigorously-determined Ld + Lo thermodynamic tieline in DSPC/POPC/cholesterol at 23°C (3). The average molecular area for each phase was determined by assigning areas to the individual lipid components in each phase, and computing the mole-fraction weighted sum of these values. For both Ld and Lo phases, the cholesterol area was set to 0.285 nm$^2$ (13,14), and the area of DSPC was set to its value of 0.475 nm$^2$ in the tilted gel phase (15). Little information is available about areas in POPC/cholesterol, so we made use of the experimental observation that DOPC and POPC are very similar structurally (16). We used a value of 0.631 nm$^2$ for POPC, reflecting the decrease in DOPC area that occurs with the addition of 10 mol% cholesterol (17). Hydrocarbon thicknesses for Ld and Lo phases of 27.1 Å and 37.1 Å (respectively) were taken from POPC and DSPC values (15,16), with 2.2 Å added to the POPC thickness to account for the cholesterol content of the Ld phase (17). No offset was applied to the Lo phase thickness, as it has been shown that the addition of cholesterol to saturated gel-phase lipids has little effect on bilayer thickness. The percolation threshold has not been determined for DSPC/POPC/cholesterol, but the threshold in similar systems appears to be approximately half-way along the tieline, so we used a value of $\chi_{Lo} = 0.5$.

*Probe parameters*

Probe parameters were based on commercially available fluorescent lipid analogs: DHE donor to DiO acceptor, and DiO donor to DiI acceptor. Donor fluorescence lifetimes were set to 1.0 ns for both donors in both phases, and relative donor fluorescence quantum yields in the Ld and Lo phases were set to 1.0. Förster distances in each phase were set to 2.5 nm for DHE to DiO and 5.7 nm for DiO to DiO (18). Acceptor mole fractions were 0.005 (DiO) and 0.001 (DiI).

The transverse positions of chromophore planes were referenced to electron density profiles of POPC and DSPC bilayers, with adjustments made for the effects of cholesterol on bilayer thickness (15-17). The DiI and DiO chromophores were taken to reside 6.3 Å below the phosphate peak (19). The chromophore position of DHE was estimated from a molecular model to be 5-8 Å below the hydroxyl, which is approximately the same transverse location as the DiI and DiO chromophores; for convenience, the same value was used.

Probe radii are required to determine the distance of closest approach between donor and acceptor. Cholesterol radius was taken to be $\sqrt{0.37 \; nm^2/\pi} = 0.34 \; nm$. Average DiI radius was calculated from MD simulations of DiI in fluid DPPC bilayers: briefly, the diameter (defined as the distance across the chromophore) was calculated for each of $10^4$ frames in a 100 ns movie and averaged.

*Simulation details*

Energy transfer simulations were performed at 51 evenly-spaced compositions along the tieline. At each composition, Monte Carlo bilayer snapshots were constructed by randomly placing non-overlapping round domains of specified radius in a square simulation box with periodic boundary conditions. The dimensions of the box were chosen based on the acceptor mole fraction so that ~1000 acceptors would be found in the box, and subject to the constraint of an integral number of domains. Two leaflets with different acceptor configurations were then constructed from each domain configuration, corresponding to the physical constraint of cross-leaflet coupling between phase domains. The average number of acceptors in the domain and surround phases was calculated from box dimensions, total acceptor mole fraction, and acceptor partition coefficients; for each leaflet snapshot, the actual number of acceptors to be placed in domains and surround was drawn from a Poisson distribution using the average values. Trial acceptor coordinates were generated and tested for location in a domain and non-overlap with existing acceptors: this process was repeated until the specified number of acceptors in each leaflet was placed.

The steady-state transfer efficiency observed in an ensemble sample is an average of transfer efficiency from donors in each phase, weighted by the relative numbers of donors in those phases. (N.B.: If $E_{FRET}$ is calculated from steady-state donor emission, the weight includes the relative quantum efficiency of the donor in each phase; for simplicity, we assumed equal quantum efficiencies in this study). For each bilayer snapshot, the ensemble $E_{FRET}$ was determined by averaging $E_{FRET}$ calculated for individually placed donors. The total number of donors to be averaged from the domain and surround phases was calculated from the total number of donors ($N_D$, typically $10^4$), the partition coefficient of donor into the domain phase, and the mole fraction of domain phase:

$$N_{D,d} = N_D K_{D,d} \chi^d / (1 - \chi^d + \chi^d K_{D,d}) \qquad 1$$

$$N_{D,s} = N_D - N_{D,d} \qquad 2$$

Trial donor coordinates were generated randomly and tested for both non-overlap with existing acceptors and inclusion in a domain, rejecting trial coordinates as needed to achieve the correct domain and surround counts. For each donor $D_j$, the set of distances $\{r\}$ to all acceptors $A_k$ within a cutoff radius $20\ R_0$ (including acceptors in both leaflets) was determined. Following (20), fluorescence decay for the j$^{th}$ donor in the i$^{th}$ phase was calculated from $\{r\}$ and $R_0$:

$$\rho_{ij,DA}(t) = e^{-t/\tau_i} \prod_{k=1}^{N_A} e^{-t/\tau_i (R_{0,i}/r_{jk})^6} \qquad 3$$

The average decay for the set of $N_D$ donors in the presence of acceptors is given by:

$$i_{DA}(t) = 1/N_D \sum_i \sum_{j=1}^{N_D} \rho_{ij,DA}(t) \qquad 4$$

The decay function for donors in the absence of acceptors is:

$$i_D(t) = 1/N_D \sum_i N_{D,i} e^{-t/\tau_i} \qquad 5$$

and the ensemble transfer efficiency is:

$$E = 1 - \int_0^\infty i_{DA}(t)dt \bigg/ \int_0^\infty i_D(t)dt \qquad 6$$

The above process was iterated and the results averaged until the desired confidence for $E_{FRET}$ was achieved.

Additional simulations were performed in which various assumptions of the FP-FRET model were relaxed; the specific details of these simulations are discussed in the Results section.

<u>Nonlinear least-squares fitting of simulated data</u>

*Domain surface coverage function*

The ensemble-averaged domain surface coverage function $\langle \sigma^d(r',f) \rangle$ (see Equation 27 in the companion paper) was obtained from Monte Carlo simulations. Briefly, a domain snapshot with periodic boundary conditions was generated for unit-radius disks at a given packing fraction $f$ as described in the previous section. The domain surface coverage function, which gives the domain density as a function of distance from an "average" point within a domain, was obtained by iterating the following procedure:

1. A random disk center $(x,y)_i$ was chosen from the set of disk centers in the snapshot.

2. A random point on the disk, $p_i = (x_i + \sqrt{r} \cos\theta, y_i + \sqrt{r} \sin\theta)$, was chosen by drawing random values from the uniform distributions $r \in [0,1]$ and $\theta \in [0,2\pi)$.

3. An additional $10^4$ random points $\{q\}$ were generated as in step 2, with the value of $r$ drawn from the uniform distribution $r \in [0, c^2]$, where $c$ is a pre-determined cutoff radius. The cutoff is chosen by balancing computational efficiency with useful information: as $r'$ increases, $\langle \sigma^d(r', f) \rangle$ oscillates about $f$ with decaying amplitude, and at some distance the amplitude of the oscillations are no longer meaningful in the calculation of transfer efficiency, which decays as $(R_0/r)^6$ (see Figure 5 B in the companion paper). We determined that for the largest value of $f$ simulated (0.65), this cutoff occurs at $r' < 14$: for all simulated packing fractions, $c$ was therefore set to 14.

4. The set of distances $\{r\}_{i,q}$ from the reference point $p_i$ to each $q_j$ was calculated.

5. The points $\{q\}$ were tested for inclusion in a domain to generate a subset $\{d\}$, with a corresponding subset of distances $\{r\}_{i,d}$

This process was repeated for ~$10^3$ random disk points $p_i$; a new domain snapshot was then generated, and the entire process repeated. In this manner, a histogram of domain surface coverage was built up. The histogram bin width was set to 0.06 to give a total of 234 bins between $r' = 0$ and $r' = 14.04$. The $\sigma^d$ value for each bin was calculated from the ratio of points found in domains (the binned values of $\{r\}_d$) to total points (the binned $\{r\}_q$), and the $r'$ value was taken to be the midpoint of the bin. The simulation was terminated when the desired confidence for each bin was achieved (typically, between $3 \times 10^5$ and $6 \times 10^5$ total counts for each bin in $\{d\}$). The histogram was then converted to a set of 234 data points $(f, r', \sigma^d)$ and smoothed with a Savitsky-Golay filter. A total of 65 packing fractions between 0.01 and 0.65 were simulated, to generate a combined 2D data set of 15,210 points. This data set constitutes a discrete approximation to the full ensemble-averaged domain surface coverage function, $\langle \sigma^d(r', f) \rangle$; a continuous, integrable functional representation of $\langle \sigma^d(r', f) \rangle$ was generated with a 1$^{st}$-order (linear) interpolation using Mathematica's built-in Interpolation function.

*Nonlinear least-squares fitting of $E_{FRET}$ data*

Simulated $E_{FRET}$ data were fit to the FP-FRET model using a custom Levenberg-Marquardt algorithm written in Mathematica (available upon request from the authors). The partial derivatives of the model with respect to the variable parameters (i.e., the Jacobian matrix *J*) were calculated with a forward-difference approximation (21). All numerical integrals were performed using Mathematica's built-in NIntegrate function with the integration method set to "Automatic", and the integration accuracy set to 10 digits of accuracy. Function calls to the model were performed in parallel on a 128-kernel Mathematica cluster, using Mathematica's built-in parallelization functions.

In most cases the variable parameters were donor $K_P$, acceptor $K_P$, and domain size $R$; these variables were searched in logarithmic space. In linear space, $K_P \in (0,1]$ and $K_P \in [1,\infty)$ represent the same range of partitioning behavior, but into opposite phases; with logarithmic scaling, these ranges become $(-\infty, 0]$ and $[0, \infty)$. Logarithmic scaling therefore produces variables that are symmetric about 0 in terms of their partitioning strength, and which have the appropriate ranges for an unconstrained nonlinear fit $(-\infty, \infty)$. The domain size $R$ has a similar logarithmic effect on $E_{FRET}$, and is also naturally constrained to positive values. We found that this simple conversion to logarithmic space dramatically improved the convergence properties of the fitting routine. Confidence limits were computed from the covariance matrix at the solution in logarithmic space, and then converted back to linear space. All parameter values and confidence intervals in the text are reported in linear space.

**RESULTS**

*Simulations of transfer efficiency*

Table 1 lists the bilayer and probe parameters used as input in Monte Carlo simulations of tieline energy transfer efficiency ($E_{FRET}$), and Figure 1 shows the system schematically. The simulated bilayer system represents an experimentally established Ld + Lo tieline in DSPC/POPC/cholesterol (3): properties of the Ld and Lo phases including average molecular areas and thicknesses were set based on the tieline endpoint compositions as described in Materials and Methods. Two donor-acceptor pairs with different $R_0$ were simulated: donor DHE to acceptor DiO ($R_0$ =2.5 nm); and donor C18:2-DiO to acceptor DiI ($R_0$ =5.7). The partition coefficient of the cholesterol analog DHE was set to 2.5, the approximate value of cholesterol $K_P$ calculated from cholesterol concentration at the tieline endpoints. C18:2-DiO $K_P$ was set to 0.1, based on values determined for the structurally similar probe C18:2-DiI in the DSPC/DOPC/chol Ld + Lo region (3).

Figure 2 plots simulated $E_{FRET}$ data along the tieline with fluid domain radius fixed at 10 nm. Figure 2 *A* shows data for DHE efficiency to a set of DiO acceptors with different partitioning behavior, ranging from strong partition into Ld phase ($K_P$ =0.1, lower curve) to strong partition into Lo phase ($K_P$ =4, upper curve). Figure 2 *B* shows a corresponding set of trajectories for the DiO to DiI pair: in this case, the upper and lower curves are for DiI $K_P$ =0.1 and 4, respectively. The data in each panel of Figure 2 can be thought of as an experiment in which a single donor species is paired with a carbocyanine acceptor series with identical chromophore but different alkyl chains.

The trajectories in Figure 2 reveal qualitative behavior similar to observations of stimulated acceptor emission (SAE) in tieline trajectories (3,22). In particular, regions of enhanced or reduced FRET efficiency (REE and RRE) are seen, depending on the relative partitioning of donor and acceptor. The REE and RRE are a direct result of dramatic changes in the distribution of donor-acceptor distances that occur in the phase-coexistence region: they are observed only when both donor and acceptor have sufficiently strong $K_P$. The necessity of this condition is demonstrated in Figure 2, for the trajectories with acceptor $K_P$ =1 (green circles). In this case, despite the presence of coexisting phases, the spatial distribution of acceptors remains completely random at all compositions and no REE or RRE is observed.

Figures 3-4 show the effect of phase domain size on the $E_{FRET}$ profile. For each probe pair, simulated $E_{FRET}$ is shown for the cases of maximum REE and RRE (i.e., the upper and lower trajectories in each panel of Figure 2), and for a range of domain radii. Increasing the domain size has an effect that is qualitatively similar to increasing the strength of probe $K_P$: as phase domains grow, the peak height (or valley depth) increases rapidly at first, ultimately approaching a limiting value (the so-called "infinite phase-separation" limit). For both probe pairs, it is difficult to distinguish by eye the 40 nm (blue) and "infinite" radius (purple) data in the REE curves (though somewhat better separation is achieved in the RRE curves), demonstrating the sensitivity limits of the experiment. Finally, we note the appearance of the percolation threshold at $\chi_{Lo} = 0.5$ as an inflection in the $E_{FRET}$ profile for smaller domain sizes. This effect, which is caused by a discontinuity in the spatial distribution of probes when the continuous phase switches from Ld to Lo, disappears in the infinite-phase separation limit.

*Simulated data fit to the FP-FRET model*

Simulated data were fit to the FP-FRET model derived in the companion paper, as described in Materials and Methods. Curves in Figures 2-4 show the recovered $E_{FRET}$ profiles. In all cases, the unknown parameters were assumed to be the probe partition coefficients and the domain size $R$. Within this framework, two fitting schemes were used. In the first, each trajectory was fit independently of the others by varying donor and acceptor $K_P$ and $R$: the best-fit $E_{FRET}$ profile for this fitting scheme is shown as a dashed line in the figures. In the second scheme, three trajectories for each donor/acceptor pair were fit globally, with a total of 5 adjustable parameters: a single donor $K_P$, three acceptor $K_P$'s, and $R$ (for global fitting, the trajectory with acceptor $K_P = 1$ was omitted). Predicted $E_{FRET}$ profiles of the global fit are shown as solid lines in the figures. In all cases, both fitting schemes yield predicted curves that match the simulated data remarkably well, as judged both by eye and by reduced chi-square (see Table 2). The minor exception is for DHE/DiO REE data in the regime of Ld domains (the right-hand portion of the upper curves in Figure 2 *A* and Figure 3): a small, systematic disagreement is present here, with the model predicting lower values than are observed in the simulated data.

Table 2 lists the best-fit parameters and 95% confidence intervals corresponding to the data in Figure 2. Several trends are apparent in these data. First, it is clear that for the single-trajectory fits, the best result for both probe pairs is obtained with trajectory 1: both the recovered domain size $R$ and probe $K_P$ are in good agreement with the simulation values, a result that is observed for the other domain sizes as well (data not shown). It may be significant that of the four trajectories, acceptor $K_P$ strength (defined as $|\log K_P|$) is strongest for trajectory 1. Trajectories 2 and 3 (in which the acceptor partitions to different phases but with identical strength) perform worse than trajectory 1 for DHE/DiO. Trajectories 1-3 perform similarly for DiO/DiI, which may reflect the comparatively better S/N of these data. For both probe pairs, trajectory 4 performs poorly as judged by the parameter confidence intervals: uniform acceptor $K_P$ essentially collapses the effects of $R$ and donor $K_P$ onto a single, straight-line trajectory, drastically reducing the information content of the data. For both probe pairs, the most accurate results are obtained by global fitting of trajectories 1-3 (the final column in Table 2).

The results of global fitting of tieline $E_{FRET}$ data simulated at 6 domain sizes are shown in Table 3. For both probe pairs, the agreement between recovered values of $R$ and simulation values is good at smaller domain sizes, with accuracy decreasing as domain sizes increases. Still, recovered $R$ are within 15% of the simulation value for domain radius up to at least $10\,R_0$. The trend for probe $K_P$ is reversed: the least accurate values occur at the smallest $R$, with continuous improvement as $R$ increases. Both of these observations are consistent with the fact that in the infinite domain size limit, all information regarding domain size is lost, and the curves are entirely controlled by donor and acceptor $K_P$. We note that the true (simulation) value of $R$ is in every case but two contained in the 95% confidence interval for recovered $R$ calculated from the parameter covariance matrix, which is a reasonable outcome for a sample of 12 fits.

For both probe pairs, the global fit produces better results than any of the individual trajectory fits: in general, recovered values are closer to the simulation values, and confidence intervals are smaller. It is instructive to examine the parameter correlation coefficients $r_{ij}$ obtained from the off-diagonal elements of the covariance matrix at the best-fit solution. Table 4 lists $r_{ij}$ for the single-trajectory fits listed in the lower half of Table 2, and corresponding to the DiO/DiI data shown in Figure 2 *B*. Strong correlations ($|r_{ij}| > 0.9$) are observed between pairs of parameters for all but trajectory 3. Parameters will exhibit strong correlations when their effect on the data is similar, and we have seen in Figures 2-4 that increasing the partition coefficients and increasing the domain size have at least superficially similar effects. It can be difficult to separate these effects in the absence of a sufficient number of data points, or when S/N is poor. In short, large $r_{ij}$ are an indication that the data do not contain enough information to support all of the variable parameters.

By combining multiple data sets in a global analysis, subtle differences in parameter effects are more likely to be distinguished. Table 5 lists the correlation matrices for DiO/DiI energy transfer for all domain sizes: the matrix for the $R_{sim}$ =10 nm data corresponds to the matrices in Table 4. The pairwise parameter correlations for $R_{sim}$ =10 are in nearly every case reduced by the global analysis, with all $r_{ij}$ less than 0.8 in magnitude. Similar trends are seen at other domain sizes.

*Simulated data fit to the FP-FRET model: test of model assumptions*

To this point, the simulation parameters have exactly followed the assumptions that were used to derive the FP-FRET model. It is important to assess how the model fares when the underlying assumptions are not strictly met. Testing every case is beyond the scope of these studies, but we can address some of the most likely scenarios. Wherever literature data can be found to suggest how a real bilayer might depart from the assumptions, we have tried to incorporate that into the simulations.

*Distribution of transverse probe locations*

Figure 5 shows simulated $E_{FRET}$ data in which the transverse height for each donor and acceptor was drawn from a normal distribution centered at $d$ (see Table 1) and with a standard deviation of 0.25 nm. For comparison, data for the case of a single transverse height (i.e., Figure 2) are reproduced in a lighter shade. Clearly, the effect of a small distribution of probe heights is within the noise of the data, indicating that the FP-FRET model can be applied in this case without modification. Best-fit parameter values for global fitting are listed in the third column of Table 6, and are, within error limits, identical to the case of a single probe height.

*Uncoupling of phase domains*

Figure 6 shows simulated data for the case of uncoupled phase domains: that is, the case where the lateral positions of domains in the upper and lower leaflets are completely uncorrelated. Here, clear distortions are observed: for both the RRE and REE and relative to the case of coupled domains (reproduced in a lighter shade), transfer efficiency moves in the direction of random probe mixing. In principle this could manifest in the recovered parameters as either a reduction in probe partitioning strength or domain size (or some combination of the two): for the two data sets examined, both effects are observed (column 4 in Table 6). In the case of uncoupled domains, fitting to a model that assumes domain coupling will lead to an apparent domain size that is smaller than the true value.

*Variation in domain size across the tieline*

Experimental evidence suggests that domain size may vary in Ld + Lo coexistence regions (23). Figure 7 shows data that were simulated with a linear variation in domain radius, ranging from 5 to 20 nm across the tieline (domain size is shown on the upper x-axis scale). For clarity, only two of the simulated trajectories are shown for each probe pair, those corresponding to the maximum and minimum acceptor partition strength. Also shown is the range of $E_{FRET}$ bounded by curves predicted for 5 and 20 nm domains. For both FRET pairs, simulated data lie closer to the 20 nm curve for most of the length of the tieline, owing to the steep increase in $E_{FRET}$ that occurs at the smallest domain sizes (see Figures 3-4). The dashed line shows the results of fitting the simulated data to the unmodified FP-FRET model: that is, a single value of domain size $R$ was varied along with the probe $K_P$ to achieve the best fit to the data. Consistent with a visual insepction of the data, the best-fit domain radius for each probe pair is closer to the maximum domain size of 20 nm. In both cases, reduced chi-square (shown in Table 6, column 5) is significantly larger than for any fits discussed up to this point, a good indication that a model fails to capture an important aspect of the underlying probe distribution.

It is straighforward to account for smooth variation in domain size in our model. We modified the FP-FRET model to account for linear variation, which added one adjustable parameter: the best-fit to this model is shown as the solid lines in Figure 7. Column 6 in Table 6 shows that the recovered domain size parameters (the endpoint values) are in good agreement with the simulation values. As expected, a significant improvement in reduced chi-square is also observed.

**DISCUSSION**

*Choice of fixed parameters*

The primary goal of this report is to assess the performance of the FP-FRET model and its nonlinear least-squares implementation: given data generated with known parameters, can those parameters be accurately recovered? We chose to simulate data with snapshots of discrete, non-overlapping, randomly placed donors and acceptors, and subject to the thermodynamic constraints of the phase-separated system. This approach allows us to independently evaluate some of the more subtle assumptions of the model. The FP-FRET model is, after all, a continuous representation of an inherently discrete system: acceptors are treated as a continuous number density with spatial variation that depends on domain size, partitioning strength, and distance from a donor. It is reasonable that such a continuous description would break down for very small domains, consisting of as few as several tens of molecules, and for which the discrete nature of a lipid lattice may strongly influence energy transfer.

For a model built on the assumption of first-order phase separation, it is essential to work on a known tieline, and the well-established Ld + Lo tieline in the ternary system DSPC/POPC/cholesterol is a convenient choice (3). Molecular areas and bilayer thicknesses can be estimated from the abundant x-ray and neutron scattering data and molecular dynamics simulations available for these lipids (15-17). We emphasize that the applicability of the domain size analysis described in this report relies on the availability of such data.

The fluorescent probes chosen should also be carefully considered. The FP-FRET model assumes a single transverse location for each chromophore (which can be different in each phase). In a real membrane, the position of the chromophore will fluctuate in time, so that the single fixed value in the model represents the mean of a distribution. The accuracy of $E_{FRET}$ predicted by the model might therefore be expected to depend both on the width of the distribution of chromophore distances, and on the sensitivity of $E_{FRET}$ to changes in chromophore transverse position (which will in turn depend on $R_0$ and bilayer thickness). In any case, the best probes will be those that have relatively stable transverse locations in the bilayer. With the exception of bilayers having a high concentration of polyunsaturated chains, cholesterol is anchored in the membrane with its long axis nearly coincident to the bilayer normal, and its polar hydroxyl group at the level of the carbonyl oxygen (24), making DHE (a structural analog of cholesterol) a good candidate for transverse stability. A recent MD study of C18:0-DiI in fluid DPPC bilayers found a relatively narrow and approximately normal vertical distribution of the DiI chromophore, with an average location 6.3 Å below the DPPC phosphate and a standard deviation of ~ 2.5 Å (19). The narrow distribution was attributed to the delocalized positive charge on the conjugated-π system: when the charge was artificially eliminated, the chromophore position exhibited significantly greater transverse fluctuations. The positively charged DiO chromophore differs from DiI by a single heteroatomic substitution; it is reasonable to assume that if the positive charge is indeed responsible for stabilizing the chromophore position just below the interface, DiO will behave similarly to DiI. Finally, and by way of motivating the next section, we emphasize that the effectiveness of a global analysis of multiple data sets is related to their interconnectedness: they must share some common adjustable parameters. The DiI and DiO fluorophores are particularly attractive in this respect, as both are commercially available with a variety of different alkyl chains that exhibit markedly different partitioning behavior between Ld and Lo phases (25). By pairing a carbocyanine acceptor series with a single donor (like the hypothetical experiments in this report), we can achieve distinctly different $E_{FRET}$ profiles, all linked by a common donor $K_P$ and $R_0$.

*Single-trajectory vs global data analysis*

The most important FP-FRET parameters are those that control the spatial distribution of probes in the membrane: these are the domain size, and the donor and acceptor partition coefficient. $K_P$ between the coexisting phases can be independently measured with a variety of techniques, including fluorescence intensity or anisotropy (26,27). The fluorescence signal arises from the very local environment of a probe molecule, and as such offers the advantage that it does not depend on domain size. There are however some potential drawbacks to these measurements. It can be difficult to achieve good signal-to-noise at the low probe concentrations desired for studies of phase behavior. A related and much more serious problem concerns the *difference* in signal observed in the two phases $\Delta S$. It can easily be shown that for real data with a finite S/N, the confidence limits on the best-fit $K_P$ diverge rapidly as $\Delta S$ approaches the S/N ratio. As discussed in the previous section, the number of probes for which reliable information about transverse location is available is already quite limited; clearly, we are in a weak position if we must also rely on a large difference in quantum yield or anisotropy in the coexisting phases for this relatively small set of probes. Rather, we would like to identify controllable experimental conditions—for example, the number of independent trajectories, and the quantity of data points in a trajectory— that allow us to recover probe $K_P$ from the $E_{FRET}$ data itself.

To this end, we performed two types of fits on the simulated data trajectories, in each case with no assumed knowledge of probe $K_P$ or domain size. First, each 51-sample trajectory was fit as an independent experiment. We then analyzed three trajectories (each with the same donor $K_P$, but different acceptor $K_P$) simultaneously in a global fit. A global fit essentially uses all of the information contained in multiple data sets that share one or more parameters. A single goodness-of-fit criterion is established by summing the squared residuals for all of the data points (28). An essential aspect of global analysis is that the relative error for each data point is known or can be estimated: the appropriate weight for each residual term is the relative error, and different data sets will in general have different errors.

A perhaps surprising result of this study is that for even a single 51-point trajectory with a single probe pair, in most cases both the probe $K_P$ and the domain size can be obtained (see Tables 2-3). The accuracy depends on the strength of probe partitioning (with larger $K_P$ leading to more accurate values), $R_0$ for the probe pair (with smaller $R_0$ showing reduced accuracy at large domain sizes), and of course S/N. The clear exception is for uniform acceptor $K_P$, in which case the trajectory contains virtually no information. As expected, a better fit (as judged by the relative difference between simulated and recovered values, and the confidence interval of the recovered parameters) is achieved in every case with a global analysis. Certainly, some of the improvement can be attributed to the larger ratio of independent data points to fit parameters: the single-trajectory analysis includes 51 data points and 3 adjustable parameters, with the global analysis adding 102 data points and only two adjustable parameters. However, some of the improvement is the result of reduced parameter correlations. Figures 2-4 demonstrate that either increasing the domain size or increasing the probe partition coefficients has similar effects on the $E_{FRET}$ lineshape. This can potentially cause strong correlations in these parameters (see Table 2), which increases their confidence intervals. When we add another trajectory with the same donor but a different acceptor, the effect is to weaken the correlations among all of the parameters (Table 3). The effects of domain size and probe partitioning are essentially decoupled, leading to improved accuracy in the recovered parameters.

*Assumptions of the FP-FRET model*

We tested the performance of the FP-FRET model for a variety of simulated data sets in which particular assumptions of the model were examined. We now discuss the results of these simulations.

*Probe height distribution.*

We used a distribution of probe transverse positions that was based on MD simulations of DiI in DPPC bilayers (19). In that study, DiI transverse position was observed to fluctuate with a standard deviation of ~ 2.5 Å. As shown in Figure 5, no significant effect was observed in the simulated data due to the distribution, and we conclude that relatively narrow distributions need not be explicitly accounted for by the model in order to obtain reliable domain size measurements. Of course, sufficiently wider, or highly asymmetric distributions could show more significant deviations than were observed in our simulations. Furthermore, an incorrect estimate of the mean position may lead to larger errors. Mean fluorophore depth can in many cases be measured by addition of an external (29) or membrane incorporated (30) fluorescence quencher. MD studies of membrane fluorophores have begun to appear in the literature (19,31-36), and such studies will undoubtedly be an important tool for characterizing vertical distributions for use in bilayer energy transfer studies, particularly for cases where experimental data cannot be easily obtained.

*Cross-leaflet domain coupling*

Experimentally, cross-leaflet coupling of phase domains in macroscopically-separated bilayers is always observed: we are aware of no published exceptions. A variety of mechanisms has been put forward to explain these observations, including chain interdigitation, cholesterol flip-flop, and electrostatic coupling (37). Collins has recently proposed that surface tension at the bilayer midplane not only greatly reduces the probability of domain uncoupling, but makes even small overhang of domains very unlikely except near a critical point, where the interfacial energy vanishes (38). Using coarse-grained MD simulations of DPPC/DOPC/chol, Risselada and Marrink found evidence for the existence of this surface tension and estimated its magnitude, concluding that domain overhang > 20 nm$^2$ is effectively suppressed (39). A recent review concluded that the largest contributor to the midplane interfacial energy is chain interdigitation (37). We conclude that the assumption of strict phase domain register in coupled leaflets is well-supported. Still some uncertainty remains, particularly for nanoscopic domains, where direct experimental verification of coupling in free-standing bilayers is impossible.

When data arising from an uncoupled system are fit to the unmodified FP-FRET model, the recovered domain size is always smaller than the true value (Table 6). Fitting to an inappropriate model in this case also decreases the quality of the fit, evidenced by the larger reduced chi-square: changes in $E_{FRET}$ due to uncoupling are qualitatively different from the effects of domain size or probe partitioning, and the effect cannot be easily absorbed into the existing parameters. Figure 6 shows that the effect is exacerbated by increased acceptor partitioning strength, and by increased $R_0$. These results are easy to understand. The inverse-sixth power dependence of $E_{FRET}$ on donor-acceptor distance dramatically reduces the frequency of quenching events at distances larger than even a few nm, which is the typical distance between probes in opposing leaflets. In general, the only cross-leaflet acceptors that contribute in a meaningful way to donor quenching are those that are almost directly opposite the donor. When domains are coupled, the time-averaged acceptor density immediately above a given donor is enhanced or reduced (depending on the direction of acceptor partitioning) relative to a random acceptor distribution. Uncoupling of the leaflets destroys this correlation, and the time-averaged value of acceptor density opposite the donor is now the same as it would be for randomly distributed acceptors. Figure 6 reveals that uncoupling is more important for long $R_0$ probe pairs, where a larger fraction of the total donor quenching occurs to acceptors in the opposite leaflet. For small $R_0$ pairs and large transverse probe distances, the effect is expected to be negligible. In either case, it is straightforward to account for this effect in the model by using the average acceptor surface density in the term for cross-leaflet energy transfer.

The simulated data reveal that uncoupling is clearly a differential effect with respect to $R_0$, more significant for larger than for smaller values. Furthermore, compared to distortions in $E_{FRET}$ due to the failure of other model assumptions, this differential effect seems to be unique. This suggests an experimental test for uncoupling of small domains. Data are collected with a variety of probe pairs covering a range of $R_0$ and fit as usual to the coupled domain FP-FRET model. If the recovered domain size decreases with increasing $R_0$, the data should be refit to an uncoupled model: if a consistent value of $R$ is obtained, uncoupling of domain leaflets may be responsible for the discrepancies in the former case.

Finally, we note that theoretical calculations raise an intriguing third possibility regarding domain coupling: that below a certain critical radius, estimated to be ~ 2 nm, the most energetically favorable configuration is the *anti-registration* of Ld and Lo domains (37). Such a scenario would seem to require nearly equal area fractions of Ld and Lo phases to significantly affect model predictions, and as such may be unimportant except near the percolation threshold.

*Variation in domain size*

There is no theoretical work of which we are aware to relate the size (or size distribution) of nanodomains to lipid composition along a tieline. Experimentally, the dependence of domain size on mixture composition has been estimated in only a handful of cases (23,40,41). Though domain size measurements in those studies are at best crude, the evidence suggests that domain size may vary across the Ld + Lo region in some mixtures. Using FRET, de Almeida et al. estimated that Lo phase domains near the Ld boundary were < 20 nm in diameter, while Ld domains near the Lo boundary were ~75-100 in diameter (23). These estimates seem large based on a comparison of FRET profiles obtained in the Ld + Lo region of DSPC/DOPC/chol and DSPC/POPC/chol (3). For simulating the effects of variable domain size, a smaller variation of 5-20 nm (domain radius) was used. Our results indicate that such a linear variation in domain size can be successfully incorporated into the model (Figure 7 and Table 6). Furthermore, the inappropriateness of a single-radius model would be easily diagnosed by the fit statistics. A good strategy might be to in all cases fit to a model with varying domain size, and then use the fit statistics to evaluate the results. Similar values and strong cross-correlation between the endpoint domain size parameters would indicate that the variation in domain size across the tieline is negligible.

*Additional considerations*

It is beyond the scope of this report to investigate every possible way in which assumptions of the FP-FRET model may fail to be met in real membranes. Still, some unaddressed and potentially significant complications merit a brief discussion here.

First, we must consider the issue of the FRET orientation factor $\kappa^2$. $\kappa^2$ is related to the relative orientations of the donor and acceptor transition dipoles, shown in Figure 8 *A*, as:

$$\kappa^2 = (\cos\theta_T - 3\cos\theta_D \cos\theta_A)^2 \qquad 7$$

and is related to the Förster distance through the relationship:

$$R_0^6 = \frac{9000 \ln 10 \, Q_D J \kappa^2}{128\pi^5 N_A n^4} = C\kappa^2 \qquad 8$$

(42). In Equation 8, the photophysical parameters (which are constant for a given probe pair in a given environment) have been folded into a single constant $C$. Much print has been devoted to the problem of the uncertainty associated with $\kappa^2$. This uncertainty arises from two sources: a lack of knowledge of the probability distribution of $\kappa^2$ (which is related to the range of allowed motions of the probes), and a lack of knowledge of the rate at which the transition dipoles sample their allowed orientations (which is related to the rate of rotational diffusion of the chromophores). In many cases, it is assumed that all orientations of the transition dipoles are allowed (the isotropic assumption), and that the entire range of orientations is sampled during the transfer time (the dynamic averaging assumption). The isotropic distribution of $\kappa^2$ has been exactly calculated (43), and is shown in Figure 8 *B*: the expectation value of $\kappa^2$ on this distribution is $\langle\kappa^2\rangle =2/3$. When both the isotropic and dynamic averaging assumptions hold, it is valid to use the average value of $\kappa^2 =2/3$ (and hence, a single value of $R_0$) to describe all probe pairs (43). We will now consider these two sources of uncertainty in the context of our experiments.

First is the possibility of a restricted range of motional freedom of the lipid fluorophore. Fluorophores attached via a flexible linker to the headgroup may more or less meet the approximation of an isotropic distribution of orientations, but fluorophores located within the highly asymmetric bilayer environment, such as those considered here, almost certainly do not. A common model for rotational diffusion of lipids in membranes is wobble-in-a-cone, where the long axis of the molecule is allowed to sample an axially symmetric subset of orientations about a mean orientation (44). The fact of restricted motion does not itself invalidate the use of a single value of $R_0$ to describe all probe pairs: a single value can be used whenever the entire range of *allowed* motions (e.g., a restricted conical distribution) is sampled during the transfer time. Such a case is by definition within the dynamic averaging regime, though the distribution of $\kappa^2$ may well be unknown. For cases where the desired information is the unknown D-A separation distance (or distance distribution), this fact is of little comfort, as there are now two unknowns ($R$ and $\langle\kappa^2\rangle$) in the transfer efficiency equation. However, this is not the case for the model presented here: consider that in the single-phase compositions at the tieline endpoints, where there are no phase domains and hence no probe partitioning, the assumption of a random probe distribution has precisely accounted for all donor-acceptor separation distances (at least, to the extent that the bilayer structural parameters and probe transverse locations are known, and we have in this work assumed that those parameters are fixed by independent experiments). If we relax the assumption that $\langle \kappa^2 \rangle = 2/3$, we now have one unknown parameter ($R_0$, via its relationship to $\kappa^2$). Stated another way, we can account for the effects of restricted motion simply by allowing $R_0$ to vary in the fit. An alternative (and conceptually equivalent) procedure is to measure $E_{FRET}$ vs acceptor concentration in the pure phases and fit the data to Equations 1-3 of the companion paper to recover $R_0$ for use in the model (41). It is worth reemphasizing that this is only strictly valid in the dynamic averaging regime.

The second source of uncertainty—the averaging regime (dynamic vs. static)—seems to be a source of confusion, and is less frequently discussed in the literature. The static averaging limit holds when the motion of the transition dipoles is much slower than the timescale of energy transfer (that is, the timescale of fluorescence decay). In this case, each D-A pair has a different value of $\kappa^2$, and the average transfer efficiency of an ensemble is the average of $E_{FRET}$ for all of the individual pairs, each with its own $R_0$ determined by Equation 8 (45). It can be shown that, given the same underlying distribution of orientation factors, the value of $E_{FRET}$ obtained in the static averaging limit is always less than the value obtained in the dynamic averaging limit (45). This is an intuitive result, because in the dynamic regime every D-A pair must sample *all* orientations during the transfer time, including those most favorable for energy transfer (43).

Rotational diffusion for fluorescent lipids in membranes can be examined with fluorescence anisotropy. The analysis typically reports two rotational correlation times: a fast component that is ascribed to rotational motions of the chromophore, and a slow component that is ascribed to the whole lipid. Experimentally determined values of the fast component for dialkyl-DiI probes in fluid phases range from 0.18-1.6 ns (46,47). Gullapalli has reported a similar value of 0.99 ns obtained from MD simulations of C18:0-DiI (19). These values are comparable to the fluorescence lifetimes of these probes, an indication that the dynamic averaging regime may not be strictly valid for DiO to DiI energy transfer. Dale and Eisinger have noted that the intermediate regime is almost impossible to model analytically (45), though they have also claimed that it can in many cases be reasonably approximated by the dynamic averaging limit (43). Though modeling the intermediate regime may prove extremely difficult, it will be important to at least assess the effects of probe rotational motions on $E_{FRET}$ with MC and MD simulations.

A second issue for consideration is that of domain shape, which in this study is assumed to be perfectly round. For bilayers exhibiting macroscopic phase coexistence, round *fluid* domains are always observed away from the critical point, though it is possible that the very low line tensions inherent in nanodomain systems will result in different shapes. It has been shown that low line tensions in the presence of competing interactions can cause phase-separated systems to undergo a shape instability; in certain interaction energy regimes, this can result in elongated domains (48). Such patterns have been observed in Ld + Lo phase coexistence regions (49, J. Wu unpublished). Transfer efficiency arising from striped domains may be difficult to model analytically, though in such cases $E_{FRET}$ can still be easily simulated with Monte Carlo techniques.

Another, related issue is a potential distribution of domain sizes. Here, experimental evidence is completely lacking, though some theoretical work suggests that nanodomains may have a very narrow size distribution (50). Towles and Dan have addressed the effects of domain polydispersity on transfer efficiency measurements, concluding that as the degree of polydispersity increases, the recovered domain size decreases, owing to a biasing effect of increased sensitivity of FRET at smaller domain sizes (51).

Finally, we note the possibility that the domains are not randomly dispersed in the membrane as assumed by the model, but rather arranged in some ordered array. For example, hexagonal arrays of Ld + Lo domains are occasionally observed in DOPC-containing mixtures (49). In such cases, it may be possible to use an expression for the 2D RDF that explicitly accounts for domain interactions. The domain surface coverage is then derived from this RDF as described in the companion paper.

Apart from geometrical considerations, a larger issue is whether or not a model of first-order phase separation is valid for the case of nanoscopic domains. Several explanations for the observation of nanodomains have been developed both theoretically and experimentally, including highly nonideal mixing (52), microemulsions (53-55), and 2D critical phenomena (4,5). This and other relevant work has been summarized in recent reviews (2,56). On the other hand, much theoretical work has shown that systems undergoing first-order phase separation can also exhibit nanometer-scale domains due to spatial modulation arising from competing energetic considerations (48). The interaction driving domains to coalesce is the domain edge energy (related to the line tension), and several potential competing interactions have been identified, including lipid intrinsic curvature and electrostatic repulsion.

Regardless of the underlying mechanism of domain formation, the FP-FRET model remains a simplistic representation of a real membrane, just as every model is, by definition, a simplification of physical reality. It is perhaps best to think of the FP-FRET model as a convenient way of approximating the spatial distribution of donor-acceptor distances that will be found whenever lipids have a strong tendency to cluster. Like any model, its success or failure should be judged based on its ability to reproduce experimental data, its predictive power, and the consistency of domain sizes it reports compared with measurements from other techniques.

*Computational considerations*

We conclude with a brief discussion of the computational considerations of the FP-FRET model. The equations of the FP-FRET model do not have simple, analytical derivatives with respect to the model parameters, yet these derivatives are required in the Levenberg-Marquardt alogorithm for nonlinear least-squares fitting, in the calculation of the Jacobian matrix. The standard approach in such cases is to use a finite-difference approximation of the derivatives to estimate the Jacobian. Finite-difference methods are computationally expensive: for the forward-difference method used in this work, each iteration of the fitting routine results in $N(M + 1)$ function calls, where $N$ is the number of data points, and $M$ is the number of variable parameters. The global analyses in this work varied 5 parameters to fit 153 data points, or a total of 918 function calls per iteration. The fits shown in Table 6 required an average of 3111 function calls: this corresponds to just over 3 iterations per fit, demonstrating the efficiency of the Levenberg-Marquardt algorithm.

Still, the sheer number of function calls poses a problem. The average computation time of a function call in the global analyses was 16 seconds (on a 2.34 GHz processor). If the function calls are performed serially, a typical analysis would take ~ 14 hours to complete. By performing the function calls in parallel, this time can be reduced to ~ 3.5 hours on a quad-core machine—a long wait to be sure, but at least within the realm of acceptability. We were fortunate to have access to a 128-node Mathematica cluster, and we found that our parallel implementation of the fitting routine scaled with an average efficiency of 83%. Operating with the full cluster, a typical simulation was completed in ~ 8 minutes.

**CONCLUSIONS**

We have shown that the FP-FRET model can provide accurate estimates of membrane domain sizes up to 20 times the Förster distance. Conservatively, with suitable fluorophores it should be possible to measure domain sizes up to 40 nm with better than 10% accuracy, and up to 100 nm with better than 20% accuracy. Though the model relies on good information about the phase separated system (i.e. the compositions and molecular areas of the coexisting phases), it is robust to some less certain but physically realistic scenarios, such as a distribution of transverse probe distances and variation in domain size across the tieline. Furthermore, we demonstrated that by simply varying $R_0$ in the fit, recovered parameter values are accurate in the face of a distribution of probe orientation factors, despite the fact that such a distribution is not explicitly accounted for by the model. For these reasons, the FP-FRET model should prove valuable for determining domain sizes in nanodomain-forming bilayer systems.

# REFERENCES

1. Lingwood, D., H. J. Kaiser, I. Levental, and K. Simons. 2009. Lipid rafts as functional heterogeneity in cell membranes. Biochemical Society Transactions 37:955-960.

2. Elson, E. L., E. Fried, J. E. Dolbow, and G. M. Genin. 2010. Phase separation in biological membranes: integration of theory and experiment. Annual Review of Biophysics 39:207-226.

3. Heberle, F. A., J. Wu, S. L. Goh, R. S. Petruzielo, and G. W. Feigenson. 2010. Comparison of Three Ternary Lipid Bilayer Mixtures: FRET and ESR Reveal Nanodomains. Biophysical Journal 99:3309-3318.

4. Veatch, S. L., O. Soubias, S. L. Keller, and K. Gawrisch. 2007. Critical fluctuations in domain-forming lipid mixtures. Proceedings of the National Academy of Sciences of the United States of America 104:17650-17655.

5. Honerkamp-Smith, A. R., P. Cicuta, M. D. Collins, S. L. Veatch, M. den Nijs, M. Schick, and S. L. Keller. 2008. Line tensions, correlation lengths, and critical exponents in lipid membranes near critical points. Biophysical Journal 95:236-246.

6. Rao, M., and S. Mayor. 2005. Use of Forster's resonance energy transfer microscopy to study lipid rafts. Biochimica et Biophysica Acta 1746:221-233.

7. Sharma, P., R. Varma, R. C. Sarasij, Ira, K. Gousset, G. Krishnamoorthy, M. Rao, and S. Mayor. 2004. Nanoscale organization of multiple GPI-anchored proteins in living cell membranes. Cell 116:577-589.

8. Goksu, E. I., J. M. Vanegas, C. D. Blanchette, W. C. Lin, and M. L. Longo. 2009. AFM for structure and dynamics of biomembranes. Biochimica et Biophysica Acta 1788:254-266.

9. Eggeling, C., C. Ringemann, R. Medda, G. Schwarzmann, K. Sandhoff, S. Polyakova, V. N. Belov, B. Hein, C. von Middendorff, A. Schonle, and S. W. Hell. 2009. Direct observation of the nanoscale dynamics of membrane lipids in a living cell. Nature 457:1159-1162.

10. Sahl, S. J., M. Leutenegger, M. Hilbert, S. W. Hell, and C. Eggeling. 2010. Fast molecular tracking maps nanoscale dynamics of plasma membrane lipids. Proceedings of the National Academy of Sciences of the United States of America 107:6829-6834.

11. Tokumasu, F., A. J. Jin, G. W. Feigenson, and J. A. Dvorak. 2003. Nanoscopic lipid domain dynamics revealed by atomic force microscopy. Biophysical Journal 84:2609-2618.

12. Roark, M., and S. E. Feller. 2008. Structure and dynamics of a fluid phase bilayer on a solid support as observed by a molecular dynamics computer simulation. Langmuir 24:12469-12473.

13. Hofsass, C., E. Lindahl, and O. Edholm. 2003. Molecular dynamics simulations of phospholipid bilayers with cholesterol. Biophysical Journal 84:2192-2206.

14. Edholm, O., and J. F. Nagle. 2005. Areas of molecules in membranes consisting of mixtures. Biophysical Journal 89:1827-1832.

15. Sun, W. J., S. Tristram-Nagle, R. M. Suter, and J. F. Nagle. 1996. Structure of gel phase saturated lecithin bilayers: temperature and chain length dependence. Biophysical Journal 71:885-891.

16. Kucerka, N., S. Tristram-Nagle, and J. F. Nagle. 2005. Structure of fully hydrated fluid phase lipid bilayers with monounsaturated chains. The Journal of Membrane Biology 208:193-202.

17. Alwarawrah, M., J. Dai, and J. Huang. 2010. A molecular view of the cholesterol condensing effect in DOPC lipid bilayers. The Journal of Physical Chemistry B 114:7516-7523.

18. Buboltz, J. T., C. Bwalya, S. Reyes, and D. Kamburov. 2007. Stern-Volmer modeling of steady-state Forster energy transfer between dilute, freely diffusing membrane-bound fluorophores. The Journal of Chemical Physics 127:215101.

19. Gullapalli, R. R., M. C. Demirel, and P. J. Butler. 2008. Molecular dynamics simulations of DiI-C18(3) in a DPPC lipid bilayer. Physical Chemistry Chemical Physics 10:3548-3560.

20. Towles, K. B., A. C. Brown, S. P. Wrenn, and N. Dan. 2007. Effect of membrane microheterogeneity and domain size on fluorescence resonance energy transfer. Biophysical Journal 93:655-667.

21. Brown, K. M., and J. E. J. Dennis. 1972. Derivative Free Analogues of the Levenberg-Marquardt and Gauss Algorithms for Nonlinear Least Squares Approximation. Numerische Mathematik 18:289-297.

22. Buboltz, J. T. 2007. Steady-state probe-partitioning fluorescence resonance energy transfer: a simple and robust tool for the study of membrane phase behavior. Physical Review E 76:021903.

23. de Almeida, R. F., L. M. Loura, A. Fedorov, and M. Prieto. 2005. Lipid rafts have different sizes depending on membrane composition: a time-resolved fluorescence resonance energy transfer study. Journal of Molecular Biology 346:1109-1120.

24. Leonard, A., C. Escrive, M. Laguerre, E. Pebay-Peyroula, W. Neri, T. Pott, J. Katsaras, and E. J. Dufourc. 2001. Location of Cholesterol in DMPC Membranes. A Comparative Study by Neutron Diffraction and Molecular Mechanics Simulation. Langmuir 17:2019-2030.

25. Baumgart, T., G. Hunt, E. R. Farkas, W. W. Webb, and G. W. Feigenson. 2007. Fluorescence probe partitioning between Lo/Ld phases in lipid membranes. Biochimica et Biophysica Acta 1768:2182-2194.

26. Heberle, F. A., J. T. Buboltz, D. Stringer, and G. W. Feigenson. 2005. Fluorescence methods to detect phase boundaries in lipid bilayer mixtures. Biochimica et Biophysica Acta 1746:186-192.

27. Silvius, J. R. 2005. Partitioning of membrane molecules between raft and non-raft domains: insights from model-membrane studies. Biochimica et Biophysica Acta 1746:193-202.

28. Press, W. H. 2000. Numerical recipes in C : the art of scientific computing. Cambridge University Press, Cambridge; New York.

29. Boldyrev, I. A., X. Zhai, M. M. Momsen, H. L. Brockman, R. E. Brown, and J. G. Molotkovsky. 2007. New BODIPY lipid probes for fluorescence studies of membranes. Journal of Lipid Research 48:1518-1532.

30. Chattopadhyay, A., and E. London. 1987. Parallax method for direct measurement of membrane penetration depth utilizing fluorescence quenching by spin-labeled phospholipids. Biochemistry 26:39-45.

31. Muddana, H., R. R. Gullapalli, E. Manias, and P. J. Butler. 2011. Atomistic simulation of lipid and DiI dynamics in membrane bilayers under tension. Physical Chemistry Chemical Physics DOI: 10.1039/C0CP00430H

32. Kyrychenko, A. 2010. A molecular dynamics model of rhodamine-labeled phospholipid incorporated into a lipid bilayer. Chemical Physics Letters 485:95-99.

33. Loura, L. M., and J. P. Ramalho. 2007. Location and dynamics of acyl chain NBD-labeled phosphatidylcholine (NBD-PC) in DPPC bilayers. A molecular dynamics and time-resolved fluorescence anisotropy study. Biochimica et Biophysica Acta 1768:467-478.

34. Loura, L. M., F. Fernandes, A. C. Fernandes, and J. P. Ramalho. 2008. Effects of fluorescent probe NBD-PC on the structure, dynamics and phase transition of DPPC. A molecular dynamics and differential scanning calorimetry study. Biochimica et Biophysica Acta 1778:491-501.

35. Loura, L. M., and J. P. Ramalho. 2009. Fluorescent membrane probes' behavior in lipid bilayers: insights from molecular dynamics simulations. Biophysical Reviews 1:141-148.

36. Loura, L. M., A. J. Carvalho, and J. P. Ramalho. 2010. Direct calculation of Forster orientation factor of membrane probes by molecular simulation. Journal of Molecular Structure 946:107-112.

37. May, S. 2009. Trans-monolayer coupling of fluid domains in lipid bilayers. Soft Matter 5:3148-3156.

38. Collins, M. D. 2008. Interleaflet coupling mechanisms in bilayers of lipids and cholesterol. Biophysical Journal 94:L32-34.

39. Risselada, H. J., and S. J. Marrink. 2008. The molecular face of lipid rafts in model membranes. Proceedings of the National Academy of Sciences of the United States of America 105:17367-17372.

40. Brown, A. C., K. B. Towles, and S. P. Wrenn. 2007. Measuring raft size as a function of membrane composition in PC-based systems: Part II--ternary systems. Langmuir 23:11188-11196.

41. Brown, A. C., K. B. Towles, and S. P. Wrenn. 2007. Measuring raft size as a function of membrane composition in PC-based systems: Part 1--binary systems. Langmuir 23:11180-11187.

42. Lakowicz, J.R. 2006. Principles of Fluorescence Spectroscopy, 3rd ed. Springer, New York.

43. Dale, R. E., J. Eisinger, and W. E. Blumberg. 1979. The orientational freedom of molecular probes. The orientation factor in intramolecular energy transfer. Biophysical Journal 26:161-193.

44. Kinosita, K., Jr., A. Ikegami, and S. Kawato. 1982. On the wobbling-in-cone analysis of fluorescence anisotropy decay. Biophysical Journal 37:461-464.

45. Dale, R. E., and J. Eisinger. 1976. Intramolecular energy transfer and molecular conformation. Proceedings of the National Academy of Sciences of the United States of America 73:271-273.

46. Ariola, F. S., D. J. Mudaliar, R. P. Walvick, and A. A. Heikal. 2006. Dynamics imaging of lipid phases and lipid-marker interactions in model biomembranes. Physical Chemistry Chemical Physics 8:4517-4529.

47. Krishna, M. M., A. Srivastava, and N. Periasamy. 2001. Rotational dynamics of surface probes in lipid vesicles. Biophysical Chemistry 90:123-133.

48. Seul, M., and D. Andelman. 1995. Domain shapes and patterns: the phenomenology of modulated phases. Science 267:476-483.

49. Baumgart, T., S. T. Hess, and W. W. Webb. 2003. Imaging coexisting fluid domains in biomembrane models coupling curvature and line tension. Nature 425:821-824.

50. Frolov, V. A., Y. A. Chizmadzhev, F. S. Cohen, and J. Zimmerberg. 2006. "Entropic traps" in the kinetics of phase separation in multicomponent membranes stabilize nanodomains. Biophysical Journal 91:189-205.

51. Towles, K. B., and N. Dan. 2007. Determination of membrane domain size by fluorescence resonance energy transfer: effects of domain polydispersity and packing. Langmuir 23:4737-4739.

52. Huang, J., and G. W. Feigenson. 1993. Monte Carlo simulation of lipid mixtures: finding phase separation. Biophysical Journal 65:1788-1794.

53. Pandit, S. A., E. Jakobsson, and H. L. Scott. 2004. Simulation of the early stages of nano-domain formation in mixed bilayers of sphingomyelin, cholesterol, and dioleylphosphatidylcholine. Biophysical Journal 87:3312-3322.

54. Mitchell, D. C., and B. J. Litman. 1998. Effect of cholesterol on molecular order and dynamics in highly polyunsaturated phospholipid bilayers. Biophysical Journal 75:896-908.

55. Nicolini, C., J. Baranski, S. Schlummer, J. Palomo, M. Lumbierres-Burgues, M. Kahms, J. Kuhlmann, S. Sanchez, E. Gratton, H. Waldmann, and R. Winter. 2006. Visualizing association of N-ras in lipid microdomains: influence of domain structure and interfacial adsorption. Journal of the American Chemical Society 128:192-201.

56. Honerkamp-Smith, A. R., S. L. Veatch, and S. L. Keller. 2009. An introduction to critical points for biophysicists; observations of compositional heterogeneity in lipid membranes. Biochimica et Biophysica Acta 1788:53-63.

# TABLES

**Table 1** Parameters used in energy transfer simulations.

| Parameter | Description | Ld value | Lo value |
|---|---|---|---|
| $\chi_{DSPC}$ | Mole fraction DSPC | 0.11 | 0.52 |
| $\chi_{POPC}$ | Mole fraction POPC | 0.8 | 0.26 |
| $\chi_{CHOL}$ | Mole fraction cholesterol | 0.09 | 0.22 |
| $a$ | Average molecular area (nm$^2$) | 0.583 | 0.45 |
| $\chi_{Lo}^{perc}$ | Percolation threshold | 0.5 | |
| $R$ | Domain radius | 2, 5, 10, 20, 40, $\infty$ | |
| $R_0^{DHE \to DiO}$ | DHE to DiO Förster distance (nm) | 2.5 | 2.5 |
| $R_0^{DiO \to DiI}$ | DiO to DiI Förster distance (nm) | 5.7 | 5.7 |
| $\tau_0$ | Donor fluorescence lifetime (ns) | 1.0 | 1.0 |
| $\varphi$ | Relative donor quantum yield | 1 | 1 |
| $r_{DHE}$ | DHE chromophore radius (nm) | 0.35 | 0.35 |
| $r_{DiO}$ | DiO chromophore radius (nm) | 0.67 | 0.67 |
| $r_{DiI}$ | DiI chromophore radius (nm) | 0.67 | 0.67 |
| $d$ | Transverse chromophore plane (nm) | 1.35 | 1.75 |
| $K_P^{DHE}$ | DHE partition coefficient | 2.5 | |
| $K_P^{DiO}$ | DiO partition coefficient | 0.1, 0.25, 4 | |
| $K_P^{DiI}$ | DiI partition coefficient | 0.1, 0.25, 1, 4 | |
| $\chi_{DiO}$ | Mole fraction DiO acceptor | 0.005 | |
| $\chi_{DiI}$ | Mole fraction DiI acceptor | 0.001 | |

**Table 2** Best-fit parameters from $R_{sim}$=10 nm data sets fit to the FP-FRET model.

| | | | DHE to DiO | | | |
|---|---|---|---|---|---|---|
| **Parameter** | **Sim. Val.** | **Traj. 1** | **Traj. 2** | **Traj. 3** | **Traj. 4** | **Global** |
| $R$ | 10 | 10 (2) | 13 (6) | 14 (9) | 7 (4) | 10.0 (9) |
| $K_P^{DHE}$ | 2.5 | 2.3 (1) | 2.3 (2) | 3 (2) | 2 (5) | 2.42 (7) |
| $K_P^{DiO,1}$ | 0.1 | 0.10 (1) | -- | -- | -- | 0.100 (6) |
| $K_P^{DiO,2}$ | 0.25 | -- | 0.27 (3) | -- | -- | 0.26 (1) |
| $K_P^{DiO,3}$ | 4 | -- | -- | 3 (3) | -- | 6 (1) |
| $K_P^{DiO,4}$ | 1 | -- | -- | -- | 2 (7) | -- |
| $\chi_{red}^2$ | -- | 0.716 | 0.924 | 1.592 | 1.727 | 1.117 |
| | | | **DiO to DiI** | | | |
| **Parameter** | **Sim. Val.** | **Traj. 1** | **Traj. 2** | **Traj. 3** | **Traj. 4** | **Global** |
| $R$ | 10 | 9.7 (5) | 9 (1) | 10.4 (5) | 10 (10) | 9.9 (2) |
| $K_P^{DiO}$ | 0.1 | 0.102 (5) | 0.11 (2) | 0.107 (6) | 0.15 (9) | 0.101 (3) |
| $K_P^{DiI,1}$ | 0.1 | 0.10 (2) | -- | -- | -- | 0.107 (7) |
| $K_P^{DiI,2}$ | 0.25 | -- | 0.23 (5) | -- | -- | 0.249 (7) |
| $K_P^{DiI,3}$ | 4 | -- | -- | 4.2 (2) | -- | 4.4 (2) |
| $K_P^{DiI,4}$ | 1 | -- | -- | -- | 1.0 (1) | -- |
| $\chi_{red}^2$ | | 0.956 | 0.859 | 1.443 | 1.112 | 1.081 |

**Table 3** Best-fit parameters from global fits to the FP-FRET model.

| | DHE to DiO | | | | | |
|---|---|---|---|---|---|---|
| Parameter | $R_{sim} = 2$ | $R_{sim} = 5$ | $R_{sim} = 10$ | $R_{sim} = 20$ | $R_{sim} = 40$ | $R_{sim} = \infty$ |
| $R$ | 2.3 (1) | 5.4 (3) | 10.0 (9) | 23 (4) | 50 (20) | 800 (--) |
| $K_P^{DHE}$ | 1.93 (8) | 2.35 (9) | 2.42 (7) | 2.42 (6) | 2.43 (5) | 2.40 (4) |
| $K_P^{DiO,1}$ | 0.13 (3) | 0.107 (8) | 0.100 (6) | 0.104 (5) | 0.106 (5) | 0.112 (5) |
| $K_P^{DiO,2}$ | 0.38 (3) | 0.28 (1) | 0.26 (1) | 0.265 (9) | 0.266 (8) | 0.272 (8) |
| $K_P^{DiO,3}$ | 11.23 (6) | 6 (1) | 6 (1) | 5.9 (8) | 6.0 (8) | 6.3 (8) |
| $\chi_{red}^2$ | 1.230 | 1.072 | 1.117 | 1.042 | 1.299 | 1.052 |
| | DiO to DiI | | | | | |
| Parameter | $R_{sim} = 2$ | $R_{sim} = 5$ | $R_{sim} = 10$ | $R_{sim} = 20$ | $R_{sim} = 40$ | $R_{sim} = \infty$ |
| $R$ | 2.00 (3) | 5.07 (8) | 9.9 (2) | 20.1 (9) | 42 (4) | 90 (20) |
| $K_P^{DiO}$ | 0.05 (3) | 0.108 (6) | 0.101 (3) | 0.102 (2) | 0.101 (2) | 0.105 (1) |
| $K_P^{DiI,1}$ | 0.16 (6) | 0.10 (1) | 0.107 (7) | 0.101 (6) | 0.099 (6) | 0.098 (6) |
| $K_P^{DiI,2}$ | 0.33 (8) | 0.24 (1) | 0.249 (7) | 0.245 (6) | 0.247 (5) | 0.249 (5) |
| $K_P^{DiI,3}$ | 3.7 (7) | 4.5 (3) | 4.4 (2) | 4.2 (1) | 4.0 (1) | 3.83 (9) |
| $\chi_{red}^2$ | 1.429 | 0.884 | 1.081 | 1.493 | 1.455 | 1.473 |

**Table 4** Correlation matrices for single-trajectory fits of DiO to DiI energy transfer, $R_{sim}$=10 nm data. Bold entries indicate strong correlations ($|r_{ij}|$ >0.9).

| **Trajectory 1** | | $K_P^{DiO}$ | $K_P^{DiI,1}$ |
|---|---|---|---|
| | $R$ | -0.706 | 0.882 |
| | $K_P^{DiO}$ | | **0.934** |
| **Trajectory 2** | | $K_P^{DiO}$ | $K_P^{DiI,2}$ |
| | $R$ | -0.871 | **0.945** |
| | $K_P^{DiO}$ | | **0.979** |
| **Trajectory 3** | | $K_P^{DiO}$ | $K_P^{DiI,3}$ |
| | $R$ | 0.711 | -0.813 |
| | $K_P^{DiO}$ | | -0.288 |
| **Trajectory 4** | | $K_P^{DiO}$ | $K_P^{DiI,4}$ |
| | $R$ | -0.390 | **-0.946** |
| | $K_P^{DiO}$ | | 0.645 |

**Table 5** Correlation matrices for global fits of DiO to DiI energy transfer. Bold entries indicate strong correlations ($|r_{ij}| > 0.9$).

| $R_{sim} = 2$ | $K_P^{DiO}$ | $K_P^{DiI,1}$ | $K_P^{DiI,2}$ | $K_P^{DiI,3}$ |
|---|---|---|---|---|
| R | -0.041 | 0.167 | 0.160 | -0.225 |
| $K_P^{DiO}$ | | **-0.97** | **-0.913** | 0.139 |
| $K_P^{DiI,1}$ | | | **0.901** | -0.162 |
| $K_P^{DiI,2}$ | | | | -0.153 |
| $R_{sim} = 5$ | $K_P^{DiO}$ | $K_P^{DiI,1}$ | $K_P^{DiI,2}$ | $K_P^{DiI,3}$ |
| R | -0.094 | 0.449 | 0.498 | -0.717 |
| $K_P^{DiO}$ | | -0.880 | -0.780 | 0.441 |
| $K_P^{DiI,1}$ | | | 0.843 | -0.637 |
| $K_P^{DiI,2}$ | | | | -0.633 |
| $R_{sim} = 10$ | $K_P^{DiO}$ | $K_P^{DiI,1}$ | $K_P^{DiI,2}$ | $K_P^{DiI,3}$ |
| R | -0.057 | 0.485 | 0.529 | -0.799 |
| $K_P^{DiO}$ | | -0.795 | -0.670 | 0.361 |
| $K_P^{DiI,1}$ | | | 0.750 | -0.631 |
| $K_P^{DiI,2}$ | | | | -0.626 |
| $R_{sim} = 20$ | $K_P^{DiO}$ | $K_P^{DiI,1}$ | $K_P^{DiI,2}$ | $K_P^{DiI,3}$ |
| R | -0.030 | 0.503 | 0.537 | -0.842 |
| $K_P^{DiO}$ | | -0.717 | -0.597 | 0.294 |
| $K_P^{DiI,1}$ | | | 0.679 | -0.613 |
| $K_P^{DiI,2}$ | | | | -0.608 |
| $R_{sim} = 40$ | $K_P^{DiO}$ | $K_P^{DiI,1}$ | $K_P^{DiI,2}$ | $K_P^{DiI,3}$ |
| R | 0.0001 | 0.507 | 0.534 | -0.861 |
| $K_P^{DiO}$ | | -0.660 | -0.550 | 0.245 |
| $K_P^{DiI,1}$ | | | 0.634 | -0.598 |
| $K_P^{DiI,2}$ | | | | -0.594 |
| $R_{sim} = \infty$ | $K_P^{DiO}$ | $K_P^{DiI,1}$ | $K_P^{DiI,2}$ | $K_P^{DiI,3}$ |
| R | 0.019 | 0.522 | 0.543 | -0.875 |
| $K_P^{DiO}$ | | -0.610 | -0.514 | 0.210 |
| $K_P^{DiI,1}$ | | | 0.608 | -0.597 |
| $K_P^{DiI,2}$ | | | | -0.594 |

**Table 6** Best-fit parameters recovered from global fits of data sets with relaxed assumptions.

| | | DHE to DiO | | | |
|---|---|---|---|---|---|
| Parameter | Sim. Val. | Probe z dist. | Uncoupling | Variable R 1 | Variable R 2 |
| $R$ | 10 | 10 (1) | 8.8 (8) | -- | -- |
| $R1$ | 5 | -- | -- | 18 (2) | 3 (2) |
| $R2$ | 20 | -- | -- | -- | 19 (5) |
| $K_P^{DHE}$ | 2.5 | 2.46 (8) | 2.34 (5) | 2.08 (7) | 2.6 (3) |
| $K_P^{DiO,1}$ | 0.1 | 0.099 (6) | 0.125 (8) | 0.096 (5) | 0.105 (5) |
| $K_P^{DiO,2}$ | 0.25 | 0.26 (1) | 0.29 (1) | 0.25 (1) | 0.27 (1) |
| $K_P^{DiO,3}$ | 4 | 5.1 (6) | 7 (1) | 6.8 (5) | 5 (1) |
| $\langle R_0 \rangle$ | 2.5 | -- | -- | -- | -- |
| $\chi_{red}^2$ | -- | 0.743 | 1.267 | 2.8523 | 1.050 |
| | | DiO to DiI | | | |
| Parameter | Sim. Val. | Probe z dist. | Uncoupling | Variable R 1 | Variable R 2 |
| $R$ | 10 | 9.9 (2) | 5.9 (1) | -- | -- |
| $R1$ | 5 | -- | -- | 17.7 (8) | 5.4 (6) |
| $R2$ | 20 | -- | -- | -- | 18.8 (8) |
| $K_P^{DiO}$ | 0.1 | 0.101 (3) | 0.120 (6) | 0.079 (2) | 0.102 (2) |
| $K_P^{DiI,1}$ | 0.1 | 0.111 (7) | 0.14 (1) | 0.167 (7) | 0.105 (8) |
| $K_P^{DiI,2}$ | 0.25 | 0.254 (7) | 0.28 (1) | 0.298 (7) | 0.247 (8) |
| $K_P^{DiI,3}$ | 4 | 4.41 (2) | 3.8 (2) | 3.04 (9) | 4.3 (2) |
| $\langle R_0 \rangle$ | 5.7 | -- | -- | -- | -- |
| $\chi_{red}^2$ | | 1.389 | 2.559 | 4.424 | 1.785 |

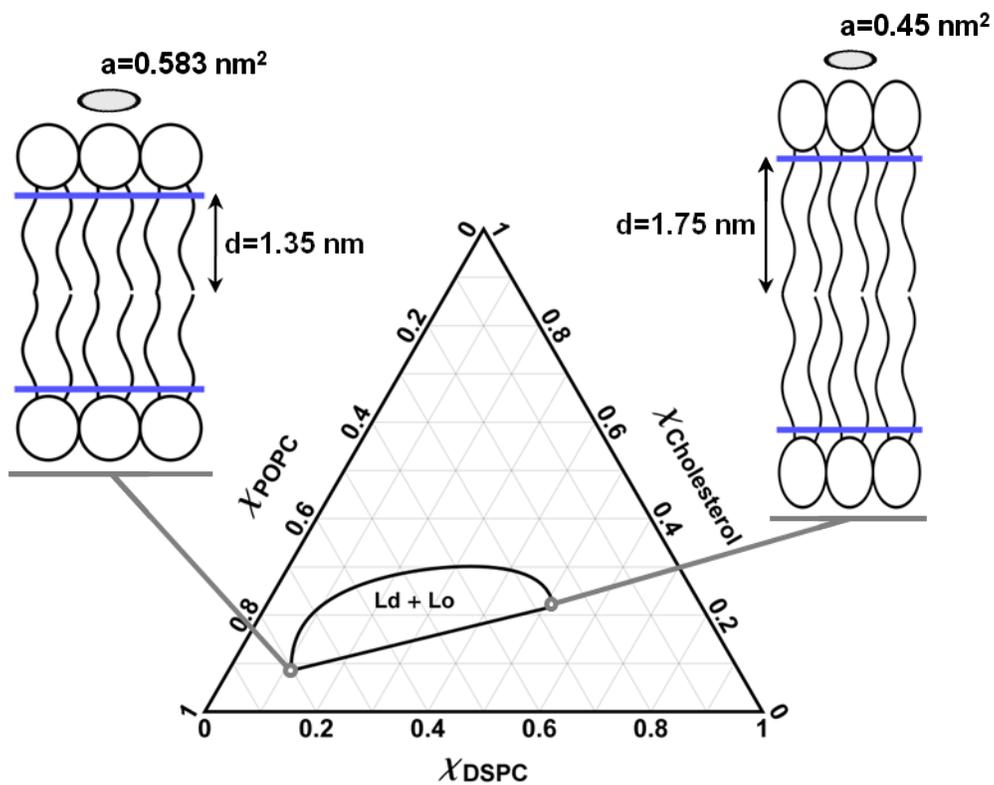

**Figure 1** Bilayer structural parameters used in nanodomain FRET simulations. Parameters are based on the Ld and Lo compositions of a tieline in DSPC/POPC/chol as described in Materials and Methods. Composition-dependent parameters include the molecular area of the phases $a$, and the mean probe transverse location $d$, measured relative to the bilayer midplane. The full list of simulation parameters is found in Table 1.

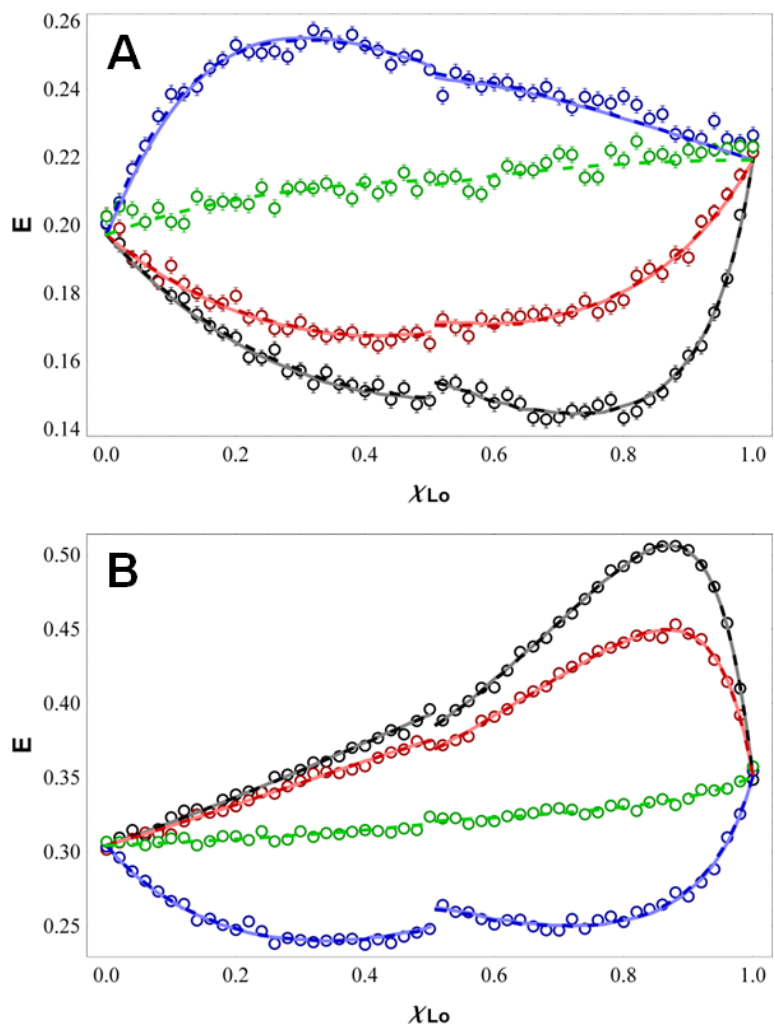

**Figure 2** Simulated FRET data reveal the effect of acceptor partitioning strength. (A) Simulated $E_{FRET}$ for DHE donor ($K_P = 2.5$) to DiO acceptor with $K_P$ 0.1 (black circles), 0.25 (red circles), 1 (green circles), and 4 (blue circles). Regions of enhanced efficiency (REE, blue circles) are observed when donor and acceptor prefer the same phase, and regions of reduced efficiency (RRE, red and black circles) are observed for opposite donor/acceptor partitioning. No change in $E_{FRET}$ (relative to a straight line joining the tieline endpoints) is observed for uniform acceptor $K_P$ (green circles). (B) Data for DiO donor to DiI acceptor with varying $K_P$ (color coding as in panel *A*). The RRE and REE are reversed relative to panel *A* due to the opposite phase preference of the donor. Fits of individual trajectories to the FP-FRET model are shown as dashed lines, and global fits are shown as solid lines.

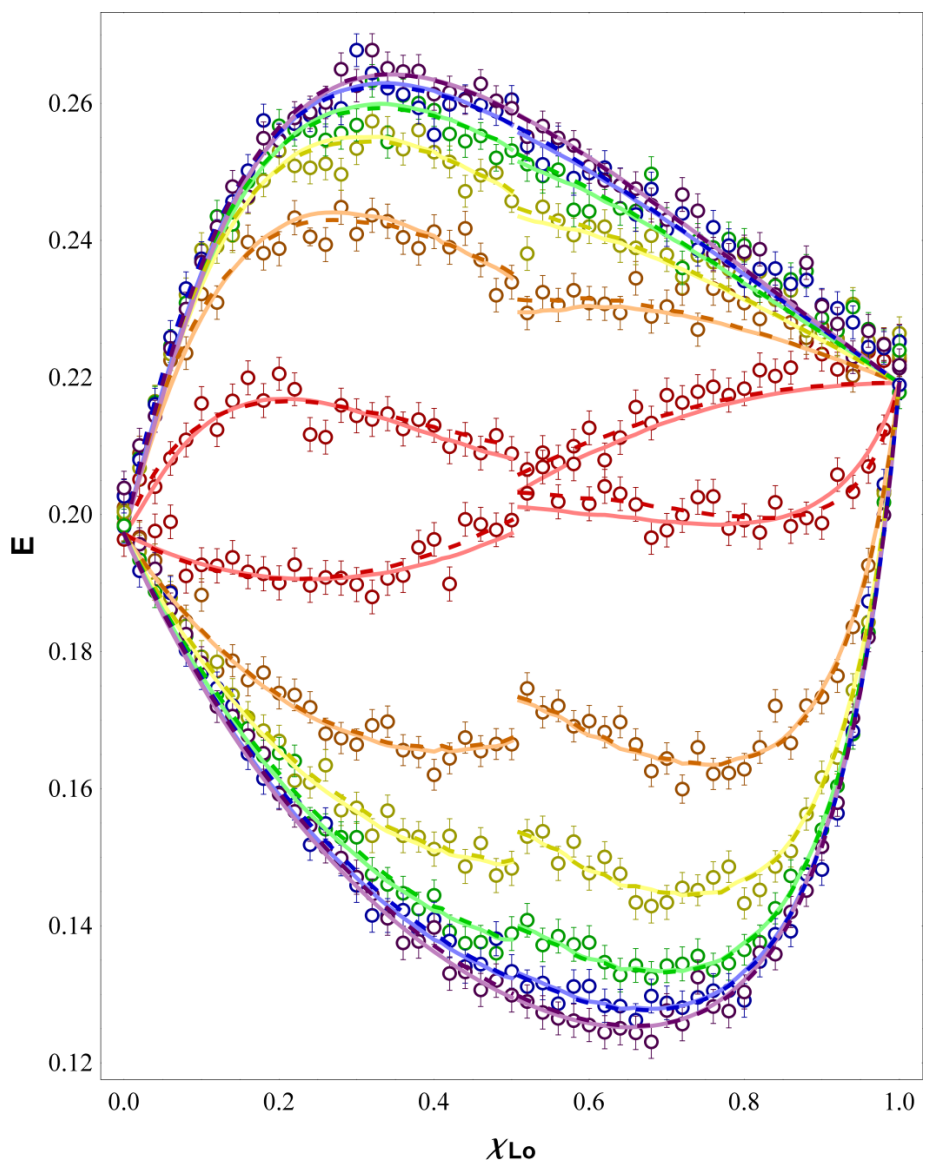

**Figure 3** Simulated FRET data reveal the effect of domain radius. Data for DHE donor to DiO acceptor with $K_P$ 0.1 (bottom 6 sets of circles) and 4 (top 6 sets of circles), for $R = 2$ nm (red), 5 nm (orange), 10 nm (yellow), 20 nm (green) and 40 nm (blue). Data predicted for the "infinite phase separation" limit, corresponding to the maximum possible domain size at each composition in a 100 nm diameter vesicle (purple). Fits of individual trajectories to the FP-FRET model are shown as dashed lines, global fits as solid lines.

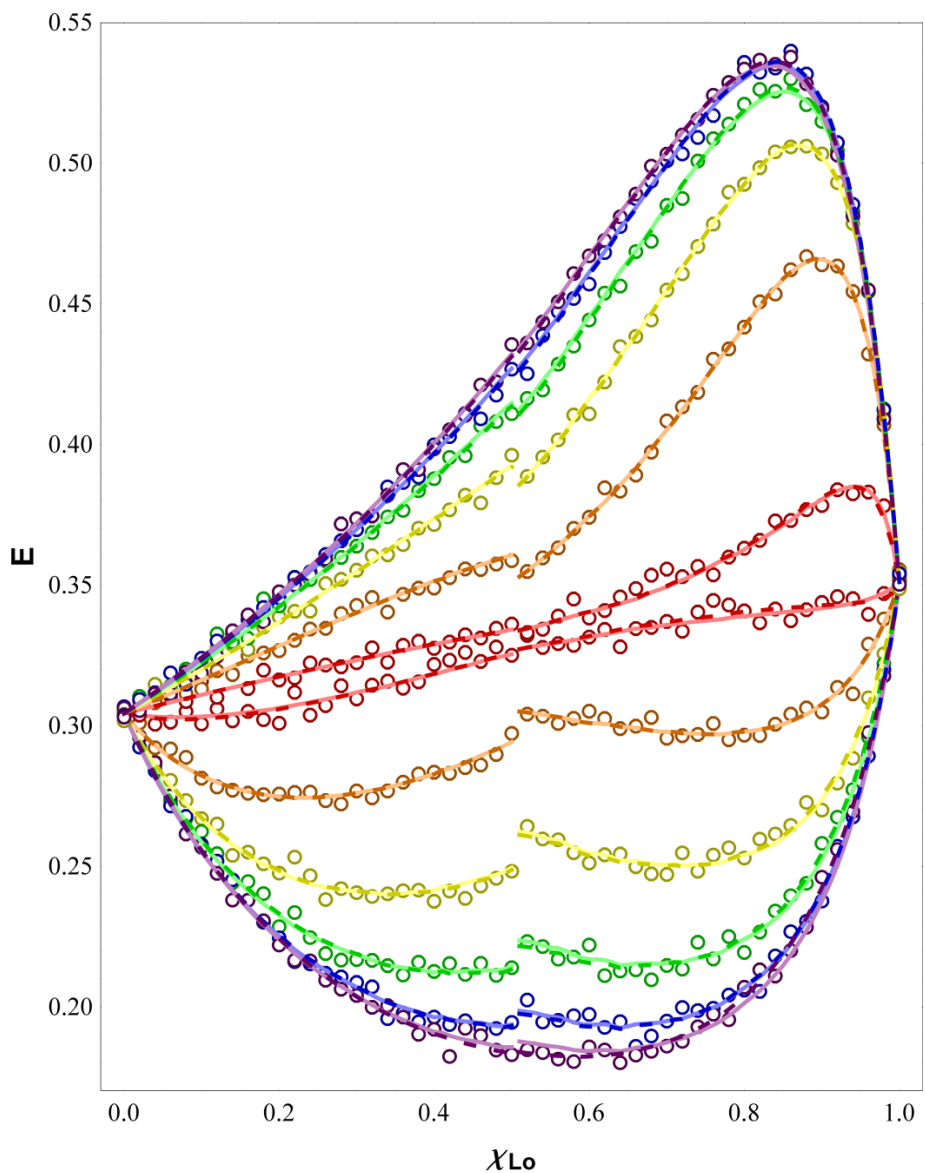

**Figure 4** Simulated FRET data reveal the effect of domain radius. Data for DiO donor ($K_P = 0.1$) to DiI acceptor with $K_P$ 0.1 (top 6 sets of circles) and 4 (bottom 6 sets of circles), for $R$ 2 nm (red), 5 nm (orange), 10 nm (yellow), 20 nm (green) and 40 nm (blue). Data predicted for the "infinite phase separation" limit, corresponding to the maximum possible domain size at each composition in a 100 nm diameter vesicle (purple). Fits of individual trajectories to the FP-FRET model are shown as dashed lines, global fits as solid lines.

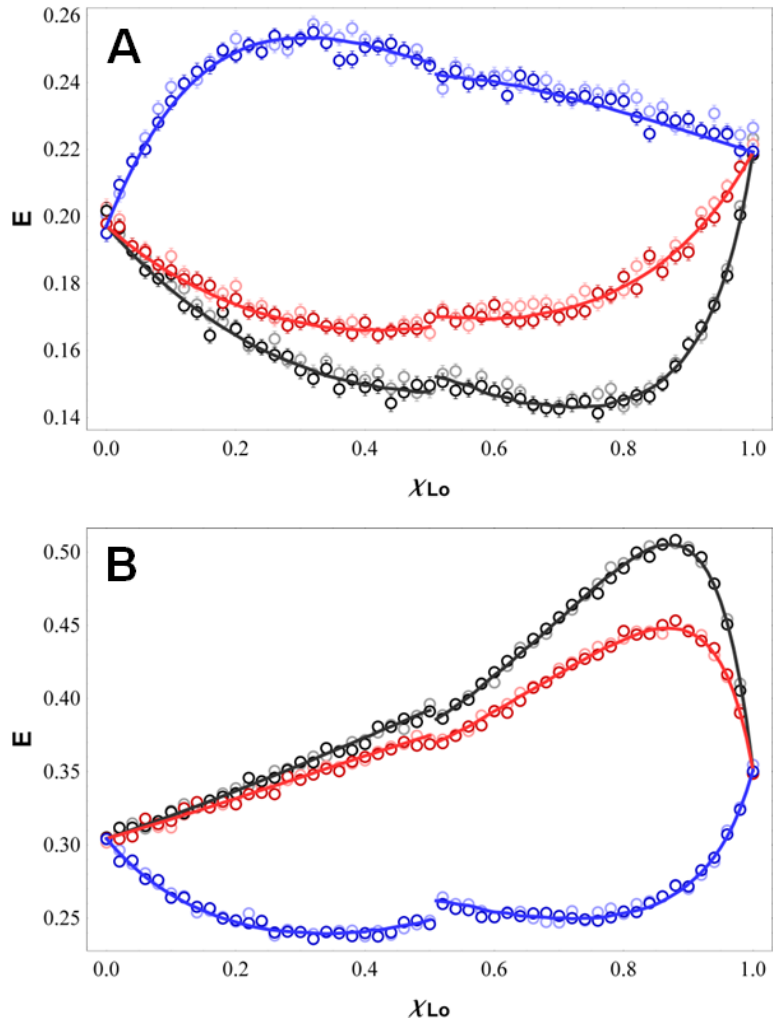

**Figure 5** A distribution of probe transverse locations has a minor effect on FRET efficiency, compared to the assumption of a single location. Simulated FRET data for DHE to DiO (A) and DiO to DiI (B), with transverse probe locations drawn from a distribution as described in the text. Colors and symbols as in Figure 2. For comparison, Figure 2 data are shown in a lighter shade. Global fits to the FP-FRET model are shown as solid lines.

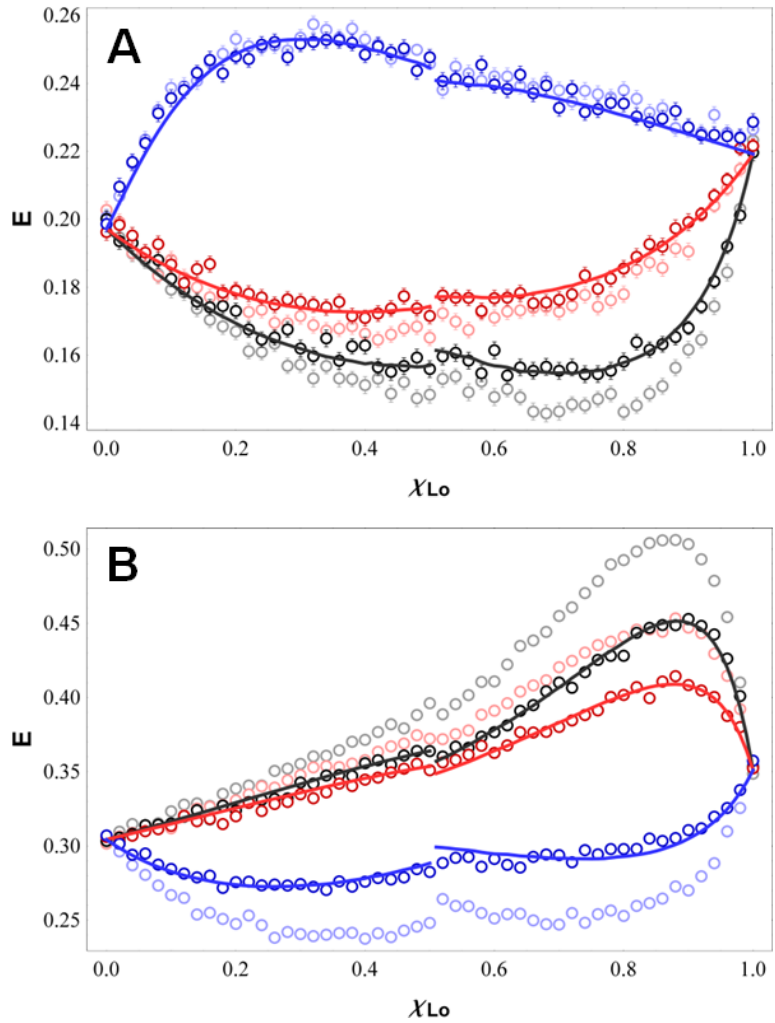

**Figure 6** Cross-leaflet uncoupling of phase domains has a large effect on FRET. Simulated FRET data for DHE to DiO (A) and DiO to DiI (B), corresponding to uncoupling of domains across leaflets as described in the text. Colors and symbols as in Figure 2. For comparison, Figure 2 data are shown in a lighter shade. Global fits to the FP-FRET model are shown as solid lines.

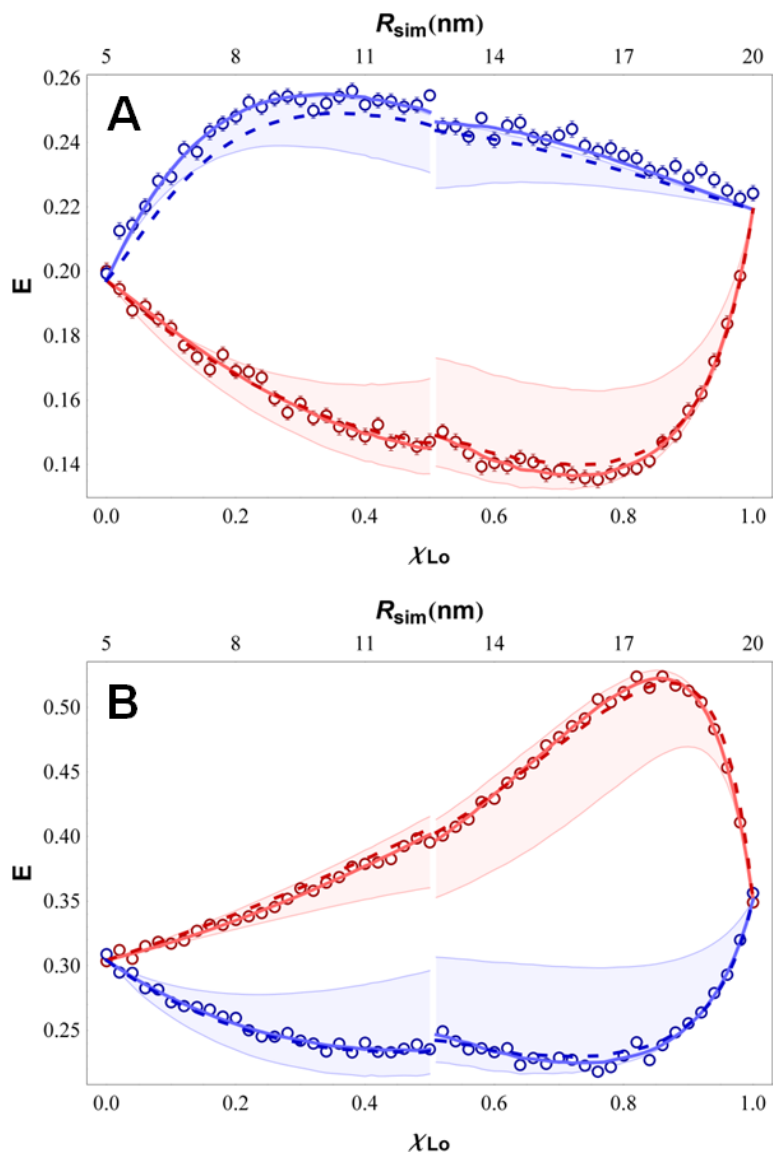

**Figure 7** Variation in domain size along a tieline affects FRET. Data for DHE to DiO (A) and DiO to DiI (B), simulated using a composition-dependent domain size shown on the upper x-axis. Colors and symbols as in Figure 2. The shaded area marks the region bounded by predicted $E_{FRET}$ values for $R = 5$ and 20 nm, for the given acceptor. Global fits to the unmodified FP-FRET model are shown as dashed lines, and global fits to a modified FP-FRET model accounting for a linear variation in domain size are shown as solid lines.

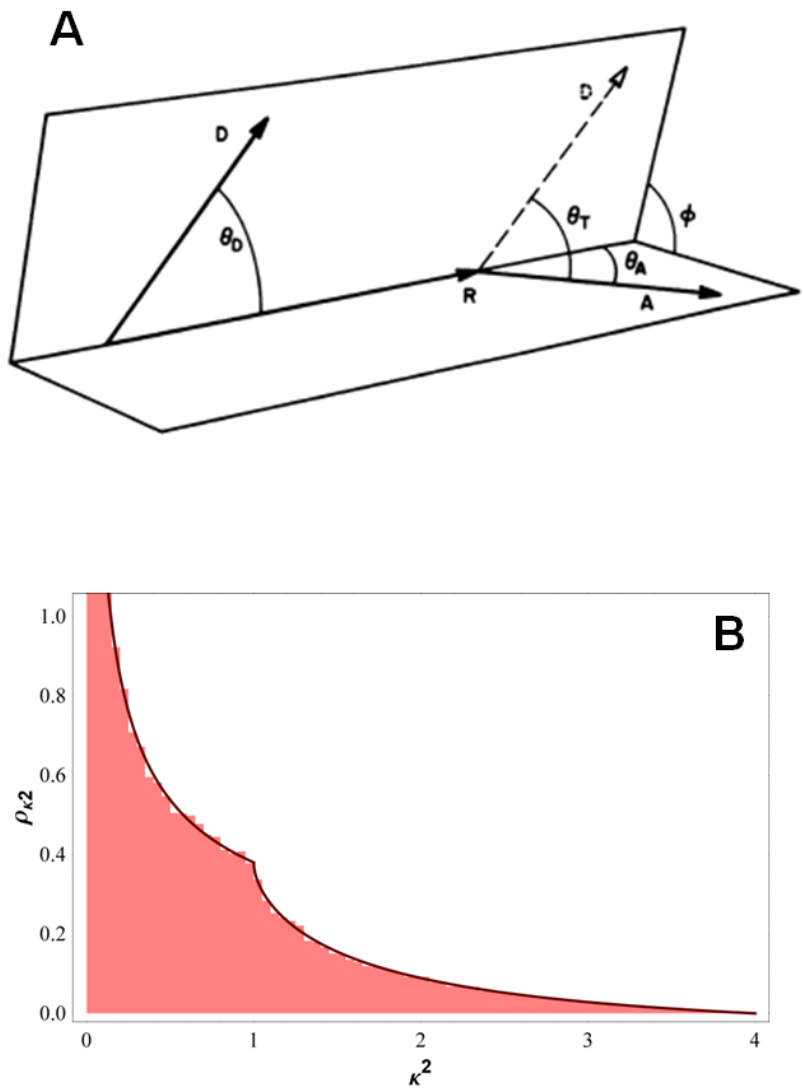

**Figure 8** FRET orientation factors are properly described by a distribution. (A) A donor excitation transition dipole D and acceptor emission transition dipole A separated by a distance R, showing the geometric considerations for calculating $\kappa^2$ with Equation 7 (from Dale et al., 1979). (B) The isotropic distribution of $\kappa^2$ resulting from complete orientational freedom of both donor and acceptor.

# APPENDIX
# A modification of the FP-FRET model to account for different bilayer thicknesses in the two phases

In this Appendix we extend the FP-FRET model developed in the companion paper for a monolayer to include the following, more physically realistic assumptions:

1. Within a monolayer, donor and acceptor are located at fixed (though potentially different) distances from the bilayer midplane.
2. In the bilayer, the phase behavior of the two monolayers is *coupled*: Domains are always in registration across the bilayer.

We start with an expression for transfer efficiency as a weighted sum of E_FRET from two donor pools, those in the domain phase and those in the surround phase:

$$E = f_D^d E^d + f_D^s E^s = f_D^d(1 - q_r^d) + f_D^s(1 - q_r^s) \qquad \text{A.1}$$

In a bilayer the quenching term is given by

$$q_r = \int_0^\infty e^{-t/\tau_0} e^{-S_{same}(t)} e^{-S_{diff}(t)} dt \qquad \text{A.2}$$

where the terms $S_{same}$ and $S_{diff}$ account for energy transfer from donors to acceptors in the same or opposing leaflets, respectively. In favorable cases the transverse location of the donor and acceptor fluorophore (that is, their position in the direction of the bilayer normal) has a narrow distribution and can be approximated by the mean. In these cases, a general form of $S(t)$ is:

$$S_{same}(t) = \int_a^\infty \langle n_A(r) \rangle 2\pi r \left[1 - e^{-(t/\tau_0)\left(R_0/\sqrt{|d-a|^2+r^2}\right)^6}\right] dr \qquad \text{A.3}$$

$$S_{diff}(t) = \int_0^\infty \langle n_A(r) \rangle 2\pi r \left[1 - e^{-(t/\tau_0)\left(R_0/\sqrt{|d+a|^2+r^2}\right)^6}\right] dr \qquad \text{A.4}$$

where $d$ and $a$ are the positions of the donor and acceptor planes with respect to bilayer midplane (see Figure 1), and $\langle n_A(r) \rangle$ is the appropriate acceptor surface density function.

Equations A.3-4 account for two different acceptor pools seen by a given donor: those in the same leaflet, and those in the opposite leaflet. If the transverse location of donor and acceptor planes is different in the coexisting phases, the number of acceptor pools doubles to four. As an example, consider the case where small domains of Lo phase exist in a continuous Ld phase. If the positions of donor planes in the two phases are given by $d^d$ and $d^s$, and that of the acceptor planes by $a^d$ and $a^s$, then donors in domains can transfer energy to acceptors at relative positions $|d^d - a^d|$, $|d^d - a^s|$, $|d^d + a^d|$, and $|d^d + a^s|$. The relative population of these pools is related to the fraction of acceptors found in domains $\langle f_A^d(r) \rangle$, given by

$$\langle f_A^d(r) \rangle = \langle \chi^d(r) \rangle K_A^d / \left[ 1 + \langle \chi^d(r) \rangle (K_A^d - 1) \right] \qquad \text{A.5}$$

where $\langle \chi^d(r) \rangle$ is the mole fraction of domain phase:

$$\langle \chi^d(r) \rangle = \langle \sigma^d(r) \rangle / a^d / \left[ \langle \sigma^d(r) \rangle / a^d + (1 - \langle \sigma^d(r) \rangle) / a^s \right] \qquad \text{A.6}$$

and $\langle \sigma^d(r) \rangle$ is the domain surface coverage, found by rearranging Equation 20 of the companion paper:

$$\langle \sigma^d(r) \rangle = \left( \langle n_A^d(r) \rangle - n_A^s \right) / (n_A^d - n_A^s) \qquad \text{A.7}$$

With these definitions,

$$\begin{aligned}
S_{same}(t) &= \int_a^\infty \langle f_A^d(r) \rangle \langle n_A(r) \rangle 2\pi r \left[ 1 - e^{-(t/\tau_0)\left(R_0/\sqrt{|d^d-a^d|^2+r^2}\right)^6} \right] dr \\
&+ \int_a^\infty (1 - \langle f_A^d(r) \rangle) \langle n_A(r) \rangle 2\pi r \left[ 1 - e^{-(t/\tau_0)\left(R_0/\sqrt{|d^d-a^s|^2+r^2}\right)^6} \right] dr
\end{aligned} \qquad \text{A.8}$$

$$S_{diff}(t)$$
$$= \int_a^\infty \langle f_A^d(r)\rangle \langle n_A(r)\rangle 2\pi r \left[1 - e^{-(t/\tau_0)\left(R_0/\sqrt{|d^d+a^d|^2+r^2}\right)^6}\right] dr$$
$$+ \int_a^\infty (1 - \langle f_A^d(r)\rangle)\langle n_A(r)\rangle 2\pi r \left[1 - e^{-(t/\tau_0)\left(R_0/\sqrt{|d^d+a^d|^2+r^2}\right)^6}\right] dr \qquad \text{A.9}$$

The ensemble-averaged functions $\langle \sigma^d(r)\rangle$, $\langle \chi^d(r)\rangle$, $\langle n_A(r)\rangle$, and $\langle f_A^d(r)\rangle$ are different for the two donor pools (those in domains, and those in the surround) as discussed in the companion paper. Finally, it should be emphasized that *all* of the ensemble-averaged functions are derived from a single radial distribution function $g(r;f)$.